\documentstyle[12pt,epsf,epsfig,pstricks,wrapfig]{article}
\begin{document}
\title{ A Generalized Spin Statistics Theorem}
{\large\noindent  \bf The Einstein-Podolsky-Rosen paradox and\\ SU(2) relativity} \vspace{5mm}

\noindent Paul O'Hara
\footnote {\small\it
Istituto Universitario Sophia,
Via San Vito, 28 - Loppiano,
50064 Figline e Incisa Valdarno (FI)
\hspace{5mm}
email: paul.ohara@sophiauniversity.org}
%
\newtheorem{thm}{Theorem}
\newtheorem{cor}{Corollary}
\newtheorem{Def}{Definition}
\newtheorem{lem}{Lemma}

\begin{abstract}
The EPR paradox dates back to 1935 when Einstein et al., through the use of non commuting operators, proposed that quantum mechanics was not complete\cite{epr} in that it suggested a ``spooky action at a distance.'' Later in 1964 John Bell was able to express the dilemma in a simple inequality\cite{bell} involving spin singlet states. If the inequality were satisfied then Einstein was correct and if it were violated then it favored the quantum mechanics point of view. In what follows, we present two approaches to Bell's inequality, and then offer an interpretation from the viewpoint of quantum mechanics based on the principle that {\it the whole is more than the sum of its parts}. This is then combined with the properties of the $SU(2)$ group to give a more intuitive understanding of non-locality and spin entanglement from the perspective of the relativistic characteristics of rotations.

\noindent KEY Words: Bell's inequality,  whole more that sum of parts, semigroup invariance, $SU(2)$ relativity.
\end{abstract}

\section{Understanding the EPR Paradox}

The original version of the EPR paradox dates back to 1935, when Einstein, Podoloski and Rosen published a paper entitled \textit{Can Quantum-Mechanical Description of Physical Reality Be Considered Complete?} The authors of the paper noted that the success of a physical theory rests on two epistemological questions: (1) Is the theory correct? and (2) Is the description given by the theory complete? They then analyzed  conjugate variables (position and momentum) from the perspective of the second question and came to the conclusion  that ``the wave function does not provide a complete description of the physical reality'' because its existence depended upon conjugacy and the measuring instrument. This contrasted with their belief  that physical realities existed in themselves and can be measured (in principle) with certainty ``{\it without disturbing the system}.''  By way of response, Niels Bohr pointed out an ambiquity in Einstein's use of the expression  `without in any way disturbing a system,' and that quantum mechanics characterizes `... an entirely new situation as regards the description of physical phenomena.''\cite{bohr}

\subsection{Bell's Inequality and the EPR Paradox}
The discussion of EPR\footnote{Nowadays,  some authors refer to this as the EPRB paradox in recognition of the fact that Bohm's reformulation of the problem in terms of the spin singlet state of the diatomic molecule was used by Bell to derive his inequality.}  might have remained merely philosophical if it had not been for the discovery of Bell's inequality in 1964 and the subsequent experimental work carried out by Aspect et al\cite{asp}. Bell began his work by interpretating EPR's belief that quantum mechanics was  incomplete to mean that the theory ``should be supplemented by additional variables''(\cite[p.~195]{bell}).  In other words, he proposed the existence of hidden variables that would conform to Einstein's notion of  measurement.  He then considered a pair of spin one-half particles in the singlet state moving away from each other in opposite directions.
Denoting their spin operators by $\bf{\sigma}_1$ and ${\bf \sigma}_2$ and their respective components of spin in the direction of the unit vector
${\bf n}$ by $\sigma_1\cdot {\bf n}$ and $\sigma_2\cdot {\bf n}$, it follows that if $\sigma_1\cdot {\bf n}$ yields a measurement of $+1$, then
by definition of a singlet state $\sigma_2\cdot {\bf n}$ will be $-1$ and vice versa. Quantum mechanics then predicts that the expected value of spin for the singlet state observed in two directions ${\bf n_1}$ and ${\bf n_2}$ will be given by 
\begin{equation}E({\bf n}_1,{\bf n}_2)=\left<\sigma_1\cdot{\bf n}_1\ \sigma_2\cdot {\bf n}_2\right>=-{\bf n}_1\cdot {\bf n}_2. \label{qmean} \end{equation}
Bell proceeded to examine the expected spin measurements in three different directions ${\bf n}_1, {\bf n}_2, {\bf n}_3$ and by means of a
triangular inequality established the relationship
\begin{equation} |E({\bf n}_1,{\bf n}_2)-E({\bf n}_1,{\bf n}_3)|\le 1+E({\bf n}_2,{\bf n}_3) \label{Bell}\end{equation}
If we now choose the three directions ${\bf n}_1, {\bf n}_2, {\bf n}_3$ in the same plane with azimuthal angles $0, \frac{\pi}{3},$ and $  \frac{2\pi}{3}$ respectively then
${\bf n}_1.{\bf n}_2={\bf n}_2.{\bf n}_3=\frac12$ and ${\bf n}_1.{\bf n}_3=-\frac12$.     
Substituting these into (\ref{Bell}) gives
$1<\frac12 $, which is a contradiction. This means that the quantum mechanics method of
calculating expectations is not compatible with the expectation derived from Einstein's et al. assumptions previously discussed.

Subsequently in 1982 Alain Aspect and colleagues experimentally measured the spin states of singlet state photons and concluded based on their experiments that nature does not obey Bell's inequalities and that the quantum expectations predictions were in agreement with their experiments. In other words, measurements associated with entangled states exhibited quantum nonlocality, where, in the words of Bell, locality meant that ``the result of measurement on one system be unaffected by operations on a distant system with which it has interacted in the past''"(\cite[p.~195]{bell}).

Generalizations of this inequality have been given by Greenberger et al. They moved beyond entangled pairs associated with Bell's inequalities to consider perfectly correlated systems of four particles \cite{green}, which means that ``knowledge of the outcomes for any three particles enables a prediction with certainty of the outcome for the fourth.'' They then established the following identity for spin observables of four perfectly correlated particles in the directions ${\bf n}_1, {\bf n}_2, {\bf n}_3, {\bf n}_4$:
\begin{eqnarray*}&&\hspace{-20 pt}  E^{\psi}({\bf n_1, n_2, n_3, n_4})=\\
&&\cos \theta_1 \cos \theta_2 \cos \theta_3 \cos \theta_4 - \sin \theta _1 \sin \theta _2 \sin \theta_3 \sin \theta_4 \cos(\phi_1+\phi_2-\phi_3 -\phi_4).\end{eqnarray*}
In the case of the perfect
 phase correlations given by $\phi_1+\phi_2-\phi_3-\phi_4=0$ with $A_{\lambda}(\phi_1)B_{\lambda}(\phi_2)C_{\lambda}(\phi_3)D_{\lambda}(\phi_4)=-1$ and
 $\phi_1+\phi_2-\phi_3-\phi_4=\pi$ with $A_{\lambda}(\phi_1)B_{\lambda}(\phi_2)C_{\lambda}(\phi_3)D_{\lambda}(\phi_4)=1$ a mathematical contradiction
 follows when it is shown that the first set of equations imply $A_{\lambda}(2\phi)=const$ for all $\phi$ while the second imply
 $A_{\lambda}(\theta + \pi)=-A_{\lambda}(\theta)$.

\subsection{Wigner's Approach}

Wigner in 1970 inspired by Bohm's gedanken experiment used a pair of spin-1/2 particles to give a simple reformulation of Bell's inequality based on the properties of the rotation group \cite{wig}. He recognized that nonlocality and the mathematical contradiction associated with the inequality followed from the group properties. Indeed, given the importance of the $SU(2)$ group in the subsequent discusion, Wigner's complete derivation is as follows:\newline
Assuming hidden parameters exist, then the set of all possible outcomes of spin values in the arbitrary chosen directions ${\bf n}_1, {\bf n}_2, {\bf n}_3$ are given by
\begin{equation}\{( +++), (++-), (+-+), (-++), (+--),( -+-),( --+), (---)\}\label{wig}\end{equation}
where $+$ indicates spin up and $-$ indicates spin down.
Specifically, if the predetermined values lie in the same plane at orientations
 $\theta_i,\ \theta_j, \ \theta_k$ with respect to a chosen axis
then

\vspace{-6mm}

\begin{eqnarray}
\;\;\;\;\;\;\{(++-), (+--)\} \subset
\{(++-),(+--),(-+-),(+-+)\} \label{wig1}\end{eqnarray}
implies
\begin{eqnarray}
P\{(++-),(+--)\} \le
P\{(++-),(+--),(-+-),(+-+)\}\label{set}.
\end{eqnarray}
Using the properties of the $SU(2)$ group and the probability interpretation associated with superposition, we obtain \begin{equation}P(++|\theta, \phi)=P(--|\theta, \phi)=\frac{1}{2}\sin^2\biggl(\frac{\theta - \phi}{2}\biggr) \label{SU1}\ .\end{equation} It then follows from equation ($\ref{set}$) that 
\begin{eqnarray}\frac 12 \sin^2 \frac {\theta_{ki}}{2}\le \frac 12 \sin^2 \frac
{\theta_{jk}}{2} + \frac 12 \sin^2 \frac {\theta_{ij}}{2}\ .\end{eqnarray}
This represents another variation on Bell's inequality. Taking
$\theta_{ij}=\theta_{jk}=\frac {\pi}{3}$ and
$\theta_{ki}=\frac {2\pi}{3}$ gives $\frac 12 \ge \frac
34$, a contradiction. The spin orientation cannot be an element of reality in the Einstein sense and measurments corresponding to
three spin states cannot be predetermined with probability 1, and in the case of entangled states measurements exhibit quantum nonlocality.

 It is also worth noting that although Wigner's derivation refers to singlet states and proves that quantum measurements involving such states are necessarily nonlocal, nevertheless equation (\ref{wig}) and the subsequent theory could also have been directly applied to a single particle. In other words, if hidden variables existed in Bell's understanding of the term then it should be possible to specify the states in all three directions in accordance with Einstein's belief. However, Wigner's inequality shows that regardless of entanglment, no such hidden variables exist (at least in the case of spin). 

\section{The EPR Paradox and Causality}

The work of Bell, Wigner and Aspect (et al.) indicate that in contrast to Einstein's belief, it is precisely the hidden variable approach that is mathematically inconsitent while quantum theory is quite complete. Nevertheless, as a vast number of publications in the area of EPR indicate, oftentimes, there is an intuitive gap between our understanding of the quantum model and our notion  of local causality. In what follows, I will suggest an alternate intuitive approach,  bearing in mind Bohr's observation that the laws of  quantum mechanics try to explain an  ``entirely new situation as regards the description of physical phenomena that the notion of complementarity aims at characterizing.'' We begin with a more complete analysis of the singlet state.
\subsection{The whole is more than the sum of its parts}

Unlike the original description of EPR proposed by Einstein, both Bohm's and Bell's approach which relie on the use of the spin singlet state, offer an additional element that can be easily overlooked and which is implicit in Bohr's reference to  ``this entirely new situation.''   We note that  in the case of the singlet state the new emerging physical laws can be subsumed under the phrase {\it the whole is more than the sum of its parts}\footnote {Aristotle, {\it Metaphysics}, London, Everyman's Library, Book Z, XI, p. 195: {\it Why the definition of a whole sometimes includes that of its parts and sometimes does not.} p. 218, VI, {\it The unity of definition: All wholes which are not mere aggregates, i.e. must have a principle of unity}}\cite{aris}. We now elaborate upon this. 

Consider the singlet state
\begin{equation}{\Psi}=\frac{1}{\sqrt 2}\left(\begin{array}{c}
                   1 \\
                   0 \\
                 \end{array}
               \right)
               \otimes \left(\begin{array}{c}
                   0 \\
                   1 \\
                 \end{array}
               \right)
-\left(\begin{array}{c}
                   0 \\
                   1 \\
                 \end{array}
               \right)\otimes
\left( \begin{array}{c}
                   1 \\
                   0 \\
                 \end{array}
               \right) \label{psi}\end{equation}
As a member of  $C^2\otimes C^2$, ${\Psi}$  is invariant under the action of the group elements $g\otimes g/|g|$ where $g\in GL(2,C)$ and $|g|$ is the determinant of $g$. Invariance means that $$|g|^{-1}(g\otimes g)\Psi = \Psi$$
In particular, if $|g|=1$ then, as has already been proven, $g\in SL(2,C)$\ \cite{IARD2018}, which also includes rotational invariance.  Moreover, as shown in \cite{IARD2018}, $\Psi$ is also  unique in that it is the only eigenstate common to all elements of $g\in SL(2,C)$.

As a tensor product, ${\Psi}$ can be embedded in $C^4$ and written as $\tilde\Psi=(0,1,-1,0)^T$ (where $T$ refers to the transpose).  Note that $\Psi = \tilde \Psi$. However, since $C^2\otimes C^2 \subset C^4$, we can say that $C^4$ is of a higher order than $C^2\otimes C^2$ and use $\Psi$ and $\tilde\Psi$ respectively to maintain the distinction  between the two spaces.  Moreover, if we consider $\tilde\Psi \in C^4$ we find that it is invariant under a 24-parameter semigroup, which we denote by $SM(4, C)$, whose elements are given by
$$A=\left(\begin{array}{cccc}
                   a_{11} & a_{12} & a_{12} & a_{14}  \\
                    a_{21} & a_{22} & a_{22}-1 & a_{24}  \\
	 a_{31} & a_{33}-1 & a_{33} & a_{34}\\
 a_{41} & a_{42} & a_{42} & a_{44}  
                 \end{array}
               \right)$$
where each $a_{ij}\in C$. Note that $a_{12}=a_{13}$, $a_{42}=a_{43}$, $a_{23}=a_{22}-1$ and $a_{32}=a_{33}-1$, which means that ${A}$ has 12 independent complex parameters. We formulate this as a lemma:
\begin{lem} $\tilde\Psi \in C^4$ is invariant under the action of the semigroup $SM(4,C)$. In other words, for all $A\in SM(4,C)$:
$A\tilde\Psi = \tilde\Psi$. 
\end{lem}
The proof follows by direct multiplication. It should also be noted that the determinant of $A$ can be any real number including 0, which also explains why $SM(4,C)$ is not a group. In general, the set of  matrices $\{A\}$ are irreducibible, although it does contain the four complex parameter reducible subgroup whose elements are  $g\otimes g/|g|$,  already mentioned above. For example, in the case of 
$$A=\left(\begin{array}{cccc}
                  1 & 0 & 0 & 0  \\
                    1 & 1 & 0 & 0  \\
	 1 & 0 & 1 & 0\\
 1 & 1 & 1 & a_{44}  
                 \end{array}
               \right) \ {\rm where} \ a_{44} \ne 1,\  A\  {\rm is\  irreducible.} $$
Whereas in the case of $ a_{44}=1$, 
$$A=\left(\begin{array}{cccc}
                  1 & 0 & 0 & 0  \\
                    1 & 1 & 0 & 0  \\
	 1 & 0 & 1 & 0\\
 1 & 1 & 1 & 1  
                 \end{array}
               \right) = \left(\begin{array}{cc}
                   1 & 0  \\
                   1 & 1  \end{array}\right) \otimes  \left(\begin{array}{cc}
                   1 & 0  \\
                   1 & 1  \end{array}\right) \ .\qquad\qquad\ \ $$

  In either case, we can say that the semigroup  $SM(4,C)$ represents a new  ``physical reality''  in Bohr's sense associated solely with the singlet state. It does not stand on its own nor can it be associated with the individual states. The irreducibility of $SM(4,C)$ is a mathematical (or a quantum mechanical) way of saying that the whole is more than the sum of its parts. It also demonstrates that Einstein ed al.'s interpretation of reality is reductionist in that it fails to conceive of higher order realities represented by $SM(4,C)$ that cannot be understood by analyzing the separate components. The incompleteness suggested by EPR is consequence of overlooking this  new ``physical reality.'' 

As an application, we can consider ionic and covalent bonds in chemistry. Both types of bonding involve spin singlet states and consequently entanglement is intrisic to our understanding of chemistry. One cannot understand water $H_2O$ by merely studying hydrogen alone or oxygen alone.  The chemistry of water is to be
understood primarily in terms of the covalent bonds between the hydrogen and oxygen atoms, bonds which in their ground state configuration form
spin-singlet states. Indeed, from the perspective of chemistry, the spin-singlet state constitutive of covalent bonding is an essential element of reality.
It represents a rotationally invariant pure state that cannot be decomposed into its individual components (collapsed wave function) without destroying
the correlation. It's a concrete example of the whole being greater than the sum of its parts.

\subsection {Entanglement and the Projection Postulate }

Keeping in mind  the previous discussion of  a``physical reality'' that  is not reducible to the sum of its parts and characterized by the semigroup $SM(4, C)$,  let us consider an alternate approach to the EPR problem.  For simplicity, we will work with those elements of reality that are specifically related to the invariant subgroup $g\otimes g/|g|$ where $g \in GL(2,C)$. Moreover,  each $g\in GL(2,C)$ can be reparameterized in terms of  polar coordinates defined with respect to a non-orthogonal frame: 
$$g=\left(
  \begin{array}{cc}
    w & y \\
    x &  z \\
  \end{array}
\right) = \left(
  \begin{array}{cc}
    r_1\cos \theta & r_2\sin \phi \\
    -r_1\sin \theta & r_2\cos \phi \\
  \end{array}
\right)$$
A frame becomes orthogonal only when dealing with proper (and improper) rotations. For convenience, we change notation and write $R(\theta,\phi)$ for $g$ to reflect the dependancy upon polar coordinates. There is also a dependancy upon $r_1$ and $r_2$. However, for the discussion that follows, we let  $r_1r_2=1$ or equivalently $|R(\theta, \phi)|=\cos(\theta-\phi)$ and so we do not explicitely refer to them in the notation. 
 
Once again, let us consider a singlet state. Three states can be associated
with it. The first is given by $\Psi$ (cf. equation (\ref{psi})). It is an invariant pure eigenstate of the operator $R(\theta,\phi)\otimes R(\theta,\phi)/| R(\theta,\phi)|$.  The second and third are reduced states (accessed through a Stern-Gerlach experiment) which come from
taking a spin measurement of the singlet state in the direction of ${\bf n}$  and are given by $\left|{\bf n},+\right>\left|{\bf n},-\right>$ and
$\left|{\bf n},-\right>\left|{\bf n},+\right>$ respectively. In terms of the elements of reality associated with EPR, we can speak of elements of reality prior
to spin measurements (the singlet state), and elements of reality after the spin measurements (the two reduced states). 

Quantum mechanics is something new. We may say that the original spin singlet state and the reduced SG states are related
as coherent and decoherent states of spin. At any one time the system is either in an entangled or a reduced state but not contemporaneously. They are mutually exclusive.
We can observe one or the other. In the case of entanglement the correlation is intrinsic to the definition
of the state and is not an epiphenomenon. The whole is greater than the sum of the parts, with the rotational invariance of the pure singlet state
constituting an additional element that cannot be captured by either of the reduced states alone and independent of any choice of angle.

Specifically, if
$${\bf a}= \left(
                                              \begin{array}{c}
                                                a_1 \\
                                                a_2 \\
                                              \end{array}
                                            \right),
                                            \ {\bf b}= \left(
                                              \begin{array}{c}
                                                b_1 \\
                                                b_2 \\
                                              \end{array}\right)$$
then the wedge product ${\bf a}\wedge {\bf b} ={\rm det}[{\bf a},{\bf b}]\sqrt{2}{\Psi}$, where $\Psi$ is the singlet state of equation (\ref{psi}) and ${\rm det}[{\bf a},{\bf b}] = a_1b_2 -b_1a_2$. Note that we use ${\bf a}$ and ${\bf b}$ to represent general vectors and reserve ${\bf n}_i$ to designate the  direction of the spin components $\sigma.{\bf n}_i$ measured in the different directions ${\bf n}_i$. This use of ${\bf n}_i$ is consistent with notation already introduced.  Consequently by  invariance  
\begin{eqnarray} &&R(\theta,\phi)\otimes R(\theta,\phi)({\bf a}\wedge {\bf b})\\
&\equiv&  R(\theta,\phi){\bf a}\wedge R(\theta,\phi){\bf b}\\
 &=&\cos(\theta-\phi)({\bf a}\wedge {\bf b}) \label{singlet}\ .
\end{eqnarray}
Moreover, although the spin operator $$\sigma_3=\left(
  \begin{array}{cc}
    1 & 0 \\
    0 & -1 \\
  \end{array}
\right)\ $$
is such that $[\sigma_3,R(\theta,\phi)]\ne 0$ and
$[\sigma_3\otimes \sigma_3,R(\theta,\phi)\otimes R(\theta,\phi)]\ne 0$ for all $\theta \ne 0$ and $\phi \ne 0$, nevertheless in the latter case the two operators share an eigenvector in common, namely
$({\bf a}\otimes {\bf b} - {\bf b}\otimes {\bf a})$.
It follows that $$[\sigma_3\otimes \sigma_3,R(\theta,\phi)\otimes R(\theta,\phi)]\ne 0\ \textrm{but}\
[\sigma_3\otimes \sigma_3,R(\theta,\phi)\otimes R(\theta, \phi)]({\bf a}\wedge {\bf b})=0$$
and
$$(R(\theta,\phi)\sigma_3\otimes R(\theta,\phi)\sigma_3)({\bf a}\wedge {\bf b})=-\cos(\theta-\phi)({\bf a}\wedge {\bf b})$$
In other words, they
commute while operating on the subspace
$$sp\{{\bf a}\otimes {\bf b}-{\bf b}\otimes {\bf a}\}\ ,$$
spanned by the rotationally invariant state ${\bf a}\wedge {\bf b}$.
It also follows that the expected value of the operator restricted to the above domain is given by
\begin{equation}\left<R(\theta,\phi)\sigma_3\otimes R(\theta,\phi)\sigma_3\right>= -\cos(\theta-\phi)= -{\bf n}_1.{\bf n}_2 \end{equation}
which is as expected from quantum mechanics. 

This offers a new insight into the nature of coherency and the projection postulate. Spin measurements of the isotropically spin-correlated states  in directions of $\theta$ and 
$\phi$, or equivalently in the directions ${\bf n}_1$ and ${\bf n}_2$ respectively,  depend only on the difference $\lambda=\theta - \phi$ and this difference is a consequence of its action on ${\bf a\wedge b}$. It reflects the isotropy of the prior correlation. The intrinsic correlation is defined prior to any projections and while in this  pure state, the singlet will continue to exhibit this correlation. This also refines our understanding of the projection postulate of quantum mechanics. A single SG measurement will destroy the isotropy and reduce the pure singlet state of the joint wave function into two independent particle states, while at the same time  maintaining a polarization in the direction of the original measurements. In other words, causality is still preserved not in a deterministic way but rather in accordance with the rules of (quantum) probabilities.   Polarization is a consequence of the SG experiment destroying the isotropic  correlation except in the direction of measurement. The polarizion is not induced by the SG magnets or a polarizing lens, rather the SG experiment destroys the entanglement except in the direction of measurement. Collapsing the wave function in this case means collapsing the isotropy.   

\subsection{Hidden Variables versus Correlations}

At the core of Bell's argument lies the fact that if hidden variables were to exist and uniquely determined the spin value of each individual particle then spin correlations 
would be merely epiphenomenal.  
This means that once we understood how the hidden parameters function with respect to each particle, we could then correlate as many particles as we wish by reproducing each one in 
the same way as prescribed by the hidden parameter. For example, to borrow from geometry, once we learn to draw an equilateral triangle with a ruler and compass, we can draw an endless number of similar equilateral triangles using the same principle.  Each equilateral triangle would then be automatically correlated. However, in this case the correlations would then be a secondary consequence of having made each triangle in a similar way. 

Returning to the singlet state, the existence of hidden parameters, in Bell's sense, would mean that it can be implicitly considered to be a product of two reduced states $A({\bf n}_1)B({\bf n}_2)$ associated
with the probability equation:
\begin{equation} E({\bf n}_1,{\bf n}_2)=-\int d(\lambda)\rho(\lambda)A({\bf n}_1,\lambda)A({\bf n}_2,\lambda) \label{bell}\end{equation}
This leads to a contradiction.  Bell's work and the subsequent experiments of Aspect et al. show that in the case of spin systems, no such hidden parameters exist. Yet, there is no doubt that in the case of a singlet state, particles are correlated. If a spin measurement is made on one particle then we immediately know the outcome of the spin measurement on the second particle, measured in the same direction. The only alternative for Bell is to seek an explanation in terms of his understanding of the projection postulate of QM.  But this too is unsatisfactory since it would mean that ``the results of one magnet now depend on the setting of the distant
magnet, which is just what we would wish to avoid'' (\cite{bell}, p.197). 

 He fails to recognize the significance of the rotational invariance of the state $\bf{a}\wedge \bf{b}$  prior to any measurements having been taken, and consequently that this can also be interpreted to mean that the pre-established
correlations are real and dependent upon a correlation parameter $\theta-\phi$, with the direction of the magnetic fields measuring this
intrinsic correlation without any need of action at a distance. In other words, the correlation $\lambda=\theta-\phi$ can itself be considered a new hidden variable but not in Bell's (nor Christian's \cite{chr}) sense, which was strictly meant to be  ``a hidden variable account of spin measurements on a single particle"(\cite[p.~196]{bell}).
Instead, here it is intrinsic to the {\it pairing}
associated with isotropically spin correlated particles. This can be seen in equation (\ref{singlet}). The $\lambda=\theta - \phi$ emerges only from the uniqueness of the wedge product ${\bf a}\wedge {\bf b}$. It cannot be extended beyond pairs and it cannot be applied to individual particles.  In other words, the spin-singlet state is a higher order phenomena that incorporates some new principle. We have already called it the ``coupling principle," meaning that {\it isotropically spin-correlated particles can only occur in pairs}\cite{ohara}. It is intrinsically related to the notation of rotational invariance or more precisely to $SM(4, {\mathcal C})$ invariance and is strictly a quantum mechanical effect. Its existence requires a new intuitive (not mathematical) understanding of the projection postulate, in that the collapsing of the wave function in this case means the collapsing of the isotropy as noted above. We explore this in greater detail below and offer an alternative interpretation in terms of $SU(2)$ relativity.\footnote{I use the words  ``$SU(2)$ relativity'' to reflect the fact that $SL(2,C)$ is homomorphic to the Lorentz group $SO(3,1)$ and that  $SU(2)\subset SL(2,C)$ is homomorphic to $SO(3)\subset SO(3,1)$. Moreover, as we shall see in sections 3.2 and 3.3, spin values are a consequence not only of the intrisic correlations related to the singlet state but also of the specific frame of reference within which it is measured.}
 
\section{Reference Frames and the SU(2) Group}
The entanglement of the singlet state guarantees that both particles will have equal and opposite values when  measured in the same reference frame. However,
entanglement does not include the assigning of unique numerical values over the entire plane which are associated not only with the correlation but
with the relative perspective of the observer. Although the SG magnets correctly observe the preexisting correlations of singlet states, Wigner's version of Bell's inequality derived  from the laws of probability and the properties of the group $SU(2)$ applied to any of the eight elements of  equation (\ref{wig}), clearly shows that such states cannot actually be consistently defined. Their existence would require that individual hidden variables be associated with each particle, in contrast with Bell's inequality which shows that such hidden variables do not exist. States such as  $(+,+,-)$  cannot be coherently defined within this context and attempting to do so hides the fact that the problem is ill-posed. 

To better understand the ill posedness, we approach the question from the perspective of what we call SU(2,C) relativity  \cite{PUL,NYAS}.
Essentially this means that although the inherent nature of the isotropically spin correlated states resides in the rotational invariance and is independent of the observer, the observed spin values are dependent upon the SU(2) frame of refence, or equivalently are dependent upon the direction of measurement of the SG instruments. In other words, the same spin state can be interpreted as $+1$ or $-1$ depending on our point of view. Consequently, as we shall see below, this permits two different values to be assigned the same event, depending on one's SU(2) frame of reference.

An analogy might help. The Earth as viewed from above the North pole can
be seen as rotating in an anti-clockwise direction (+1), while the
same state of motion can be viewed as being in a clockwise direction
(-1), if viewed from the South pole. Denoting the corresponding states
by $(1,0)$ and $(0,1)$ respectively,  if we attempt to characterize these
two states from the perspective of
at least two other frames, pointing in different directions, a contradiction
will arise if the SU(2) relativity effects are ignored. For example, we cannot consistently characterize these states from a frame on the equator. One cannot
consistently define clockwise and counterclockwise unless handedness (trivector) is taken into consideration.

In EPR something similar is happening. Each observer can uniquely describe spin up and spin down from the perspective of his
or her reference frame. However, three different observers cannot consistently define spin-up and spin-down relative to each other for a pure quantum state in terms of reduced states. The
resulting inconsistency is encapsulated in Bell's inequality.
The same reality will take on two different spin values, depending on
one's viewpoint. Moreover, since the essence of Bell's inequality involves
correlations from three different directions, its a prime candidate
for bringing forth inconsistencies, unless this SU(2) relativity
effect (which is explained below)  is taken into account.

\subsection{What is a Pure Quantum State?} In terms
of mathematics we define a pure state as an eigenvector in a Hilbert space that corresponds to an operator. The simplest example can be seen by considering 
\begin{equation}
    \left(
            \begin{array}{cc}
              1 & 0 \\
              0 &-1 \\
            \end{array}
          \right)
 \left(
 \begin{array}{c}
    1 \\
    0 \\
  \end{array}
 \right)= 
\left(
  \begin{array}{c}
   1 \\
    0 \\
  \end{array}
  \right)
\qquad
 \left(
            \begin{array}{cc}
              1 & 0 \\
              0 &-1 \\
            \end{array}
          \right)
 \left(
 \begin{array}{c}
    0 \\
    1 \\
  \end{array}
 \right)= - 
\left(
  \begin{array}{c}
   0 \\
   1 \\
  \end{array}
  \right) \label{pure}
  \end{equation}
In this case both $\left(
  \begin{array}{c}
   1 \\
    0 \\
  \end{array}
  \right)$ and $\left(
  \begin{array}{c}
   0 \\
   1 \\
  \end{array}
  \right)$
are pure states of the operator $\sigma_3$.
Now consider equation (\ref{pure}) multiplied on the left by a rotation matrix $R(\theta)$ and rewritten in the form
\begin{equation} R(\theta)\sigma_3R(-\theta)R(\theta)\left(
  \begin{array}{c}
   1 \\
    0 \\
  \end{array}
  \right)=R(\theta)\left(
  \begin{array}{c}
   1 \\
    0 \\
  \end{array}
  \right)
\end{equation}
This simplifies to
\begin{equation}  \left(
            \begin{array}{cc}
\cos 2\theta & \sin 2\theta \\
              \sin 2\theta &-\cos 2\theta \\
            \end{array}
          \right)
 \left(
 \begin{array}{c}
    \cos \theta \\
    \sin \theta \\
  \end{array}
 \right)= 
\left(
  \begin{array}{c}
   \cos \theta \\
    \sin \theta \\
  \end{array}
  \right)
\end{equation}
In other words, the  $\left(
  \begin{array}{c}
   1 \\
    0 \\
  \end{array}
  \right)$
and
$ \left(
 \begin{array}{c}
    \cos \theta \\
    \sin \theta \\
  \end{array}
 \right)$
can be viewed as the {\bf same} state but viewed from the perspective of two different reference frames and from the perspective of the different spin operators, $\sigma_3$ and $R(\theta)\sigma_3R(-\theta)$ respectively.  In this sense both are pure states.
In contrast, from the perspective of $R(\theta)\sigma_3R(-\theta)$ and $\sigma_3$ respectively they are both mixed states. It should also be noted that the spin-singlet state is an eigenvector of all linear operators defined on the tensor product of the two two-dimensional Hilbert spaces ${\cal H}\otimes {\cal H}$. With this in mind, we now introduce the notion of
$SU(2)$ relativity.

\subsection{SU(2) Relativity}

So, what is SU(2) relativity? To begin, let us erect two orthogonal frames, the laboratory frame $(i,j,k)$ and a spinor
frame $(e_i,e_{r(\theta)},e_{\theta})$, where $e_i=i$ represents
the direction of motion of a particle with spin, $e_{r(\theta)}$ is related to the direction of the detector orientated at an angle $\theta$
with respect to $i$, and $e_{\theta}=e_i \times e_r $ is a bivector. Note that relative to the laboratory frame $(i,j,k)$ there are an infinite number of
choices for $e_{r(\theta)}$ and consequently $e_{\theta}$ (one for each $\theta$). In general, for a rotation through an angle $\theta$ in the $j-k$
plane, we will obtain
\begin{equation} (e_i,e_r,e_{\theta})=(i,j\cos(c\theta)+k\sin(c\theta),k\cos(c\theta)-j\sin(c\theta))\ ,\label{triad}\end{equation}
where $c$ is a constant to be determined from the physics of the situation.

For example, in the case of a photon $c=1$. It follows that if the initial alignment of the polarizer is along the positive $j$-axis then $(e_i,e_{r(0)},e_0)=(i,j,k)$
 and  $(e_i, e_{r(\pi/2)},e_{\pi/2})=(i,k,-j)$ if the polarizer is rotated through an angle $\pi/2$ in the $j-k$ plane. This means that if
$k=e_0=\left|+\right>$ denotes spin-up in the direction $k$ then $e_{\pi/2}=-j=\left|-\right>$ denotes spin-down in the direction $-j$ with respect to
the laboratory frame.

In the case of an electron $c=1/2$, we identify $e_{r(\theta)}$ not with the direction of
the SG magnetic field but with the axes of rotation (or reflection) of the original $e_0$ after it is rotated through an angle $\theta$.
This now means that if spin-up is
represented by the vector $k$, then spin-down should be represented by a bivector $j$ with respect to the triad $(i,j,k)$. Moreover,
again based on the above, we know that if the original spin is measured as $+1/2$ with respect to
the $(i,j,k)$ triad then after a rotation of $\pi$ in the $j-k$ plane, spin will be measured as $-1/2$ in the same triad.
All of this can be encapsulated in the formula,
\begin{equation} (e_i,e_r,e_{\theta})=(i,j\cos(\theta/2)+k\sin(\theta/2),k\cos(\theta/2) -j\sin(\theta/2).\label{triad}\end{equation} It is
important to recognize that this formula reflects the physics associated with the SU(2) representation of spin. Note that for $\theta =\pi$, the spin
vector $e_{\theta}$ will be spin down and in a direction orthogonal to the original $k$.

Different frames $(e_i,e_r,e_{\theta})$ for different $\theta$ can be obtained by rotating the original triad $(e_i,e_{r(0)},e_0)$ about
different axes of rotation. However, when three frames are introduced, EPR tells us that contradictions will arise (Bell's inequality) if
we identify the results of one frame with those of another. This should not be seen as a flaw in quantum mechanics but rather as a geometrical
consequence of introducing three or more frames of the type $(e_i, e_r, e_{\theta})$ for different angles $\theta$.
In particular, consider the three frames $(e_i,e_{r(0)},e_0)$,
$(e_i,e_{r(\theta)},e_{\theta})$ and $(e_i,e_{r(\theta + \phi)},e_{\theta + \phi})$ with the understanding that the triad $(i,j,k)=(e_i,e_r,e_0)$.
It follows from (\ref{triad}) that with respect to the frame $(i,j,k)$, $$(e_i,e_{r(\theta)},e_\theta)=(e_i, j\cos (\theta/2) + k\sin (\theta/2),k\cos (\theta/2) -
j\sin (\theta/2) ) \ ,$$ and
\begin{eqnarray*}&&\hspace{-25pt} (e_i,e_{r(\theta + \phi)},e_{\theta + \phi})=\\
&&\hspace{-10pt}(e_i, j\cos ((\theta+\phi)/2)  + k\sin ((\theta + \phi)/2),k\cos ((\theta + \phi)/2)
- j\sin ((\theta + \phi)/2))\end{eqnarray*}
but with respect to the frame $(e_i,e_{r(\theta)},e_{\phi})$
$$(e_i,e_{r(\theta+\phi)},e_{\theta + \phi})=(e_i,\cos(\phi/2) e_{r(\theta)} + \sin (\phi/2) e_{\theta},\cos (\phi/2) e_{\theta} -
\sin (\phi/2) e_{r(\theta)})\ .$$
From the perspective of quantum mechanics, we can interpret this last equation to mean that the spin state $e_{\theta+\phi}$ can be identified
with the spin state $e_{\theta}$ with probability
$\cos^2(\phi/2)$ and with the spin state $e_{r(\theta)}$ with probability $\sin^2(\phi/2)$ with respect to the frame
$(e_i,e_{r(\theta)},e_{\theta})$. However, it should be noted that
$e_{\theta}= k\cos (\theta/2)  -j\sin(\theta/2)$ and therefore cannot be interpreted as being in the spin up state $k$ nor the spin down state $-j$.
Instead, $e_{\theta}$ can be interpreted as
been in the spin-up state with respect to the $(e_i,e_{r(\theta)},e_{\theta})$ but not with respect to the basis $(i,j,k)$. To ignore this leads to the
EPR paradox. A comparable situation occurs in differential geometry if one attempts to erect an orthogonal frame which uniquely spans the entire manifold. It cannot be done in general.

\subsection {Different Representations} The first thing to grasp is that
there are two equal but different representations of spin.
Specifically, consider the following:
\begin{equation}
    \left(
            \begin{array}{cc}
              1 & 0 \\
              0 &-1 \\
            \end{array}
          \right)
 \left(
 \begin{array}{c}
    1 \\
    0 \\
  \end{array}
 \right)=
\left(
            \begin{array}{cc}
             -1 & 0 \\
              0 & 1 \\
            \end{array}
          \right)
\left(
  \begin{array}{c}
   -1 \\
    0 \\
  \end{array}
  \right)
  \end{equation}
which means that \begin{equation}\left(
  \begin{array}{c}
   1 \\
    0 \\
  \end{array}
  \right)
  =-\left(
  \begin{array}{c}
   -1 \\
    0 \\
  \end{array}
  \right).
  \end{equation}

It is clear that if the ket $\left|s\right>$ represents the spin
state and $U\in SU(2)$ then $\left<s|U^*U|s\right>$ is invariant
regardless of the representation. This is what we mean by $SU(2)$
spin invariance. We now separately analyze each representation.
Returning to the notation of Greenberger et al \cite{green}, we can
write $\left|\bf n,+ \right>$ and $\left|\bf n,- \right>$ to
represent spin-up and spin-down respectively along the $\bf n$
direction. Therefore,
\begin{equation}\left|{\bf n}_1,+\right>=(\cos
\theta/2)\left|{\bf n}_2,+\right> +(\sin \theta/2)\left|\bf{
n}_2,-\right>.\label{rot} \end{equation} Now choose ${\bf n}_2$ by rotating
through $\theta=\pi/2$ (clockwise) with respect to ${\bf n}_1$ and
choose ${\bf n}_3$ by rotating through $\theta=-\pi/2$
(anti-clockwise) with respect to ${\bf n}_1$ (Fig. 1). Then equation
(\ref{rot}) gives

\begin{eqnarray} \left|{\bf n}_1,+\right>&=&1/\sqrt 2\left|{\bf n}_2, +\right>+
1/\sqrt 2\left|{\bf n}_2,-\right> \label{rot1}\\
&=&1/\sqrt 2\left|{\bf n}_3, +\right>- 1/\sqrt 2\left|{\bf n}_3,-\right>.\label{rot2}
\end{eqnarray}

\definecolor{grays}{gray}{0.90}
\begin{minipage}{0.40\textwidth}
\begin{pspicture}(-1,-2.5)(5,2.5)
\ \ \ Vector arrows$\  \ $
\psline[linecolor=gray,linestyle=dashed,  arrows=->](0,0)(4,0)
\rput(4.6,0){$\left|{\bf n}_1, +\right>$}
\psline[linecolor=red,linestyle=dashed, arrows=->](0,0)(2,2)
\rput(.4,1.2){$\left|{\bf n}_2, +\right>$}
\psline[linecolor=green,linestyle=dashed, arrows=->](2,2)(4,0)
\rput(3.6,1.2){$\left|{\bf n}_2, -\right>$}
\psline[linecolor=green,linestyle=dashed, arrows=->](0,0)(2,-2)
\rput(.4,-1.2){$\left|{\bf n}_3, +\right>$}
\psline[linecolor=red,linestyle=dashed, arrows=<-](2,-2)(4,0)
\rput(3.6,-1.2){$\left|{\bf n}_3, -\right>$}
\rput(7.5,0){{\small Fig. 1}}
\end{pspicture}
\end{minipage}

In particular if $\left|{\bf n}_1,+\right> = \left(
                                              \begin{array}{c}
                                                1 \\
                                                0 \\
                                              \end{array}
                                            \right)$,
then $\left|{\bf n}_2,+\right>=\left(\begin{array}{c} 1/\sqrt
2\\1/\sqrt 2\\
\end{array}\right)$ and $\left|{\bf n}_2,-\right>=\left(\begin{array}{c} 1/\sqrt
2\\-1/\sqrt 2\\
\end{array}\right)$, and $\left|{\bf n}_3,+\right>=\left(\begin{array}{c} 1/\sqrt
2\\-1/\sqrt 2\\
\end{array}
\right)$ and
$\left|{\bf n}_3,-\right>=\left(\begin{array}{c}-1/\sqrt 2\\-1/\sqrt 2\\
\end{array}\right)$. Note, immediately that
$\left<{\bf n}_2,\pm|{\bf n}_3,\pm \right>=0$ and
$\left|{\bf n}_2,-\right>=\left|{\bf n}_3,+\right>$. This creates the ambivalent
situation of identifying, from the perspective of ${\bf n}_1$,  a
spin-down state ($\left|{\bf n}_2,-\right>$) with a spin-up state
($\left|\bf{n}_3,+\right>$), which forces the obvious question, as to
what is the meaning of up or down in this case. In fact from the
perspective of ${\bf n}_1$ alone there is no consistent meaning; for
the value depends not only on the direction ${\bf n}_1$ but on the
choice of angle $\theta$ defined relative to ${\bf n}_1$. It is
precisely this ambivalence that is at the heart of both the  Greenberger, Horne, and  Zeilinger (GHZ)
inconsistency discussed in section 1.1  and Bell's inequality; inconsistancies that only occur if we assume the existence of hidden variables as a parameter that determines spin values independently of a reference frame. 

Finally, one might still be wondering why we cannot deduce spin values in the different reference frames with equations analogous to the Lorentz transformations of special relativity. In fact they are very similar. Once initial values of parameters and variables are specified in a given frame (usually the rest frame), Lorentz transformations allow us to calculate the values of parameters and variables such as mass,  length  and velocity in all inertial frames. Similarly, in the case of EPR, the $SU(2)$ transformations apply to probability states and once the initial probability state is specified as $(+1,0)$ or $(0,-1)$ the remaining probability states can be predicted by way of the $SU(2)$ transformations. 

\section {Conclusion} In this paper we have traced the development of the EPR paradox and have offered a new approach to understanding it based on the fact that ${\bf a}\wedge {\bf b} ={\rm det}[{\bf a},{\bf b}]\sqrt{2}{\Psi}$ is $SL(2,C)$ invariant\cite{IARD2018} and that assigned spin values depend on the $SU(2)$ frame of reference.

At the core of the EPR paradox is a new reality that comes from the fact the the whole is more than the sum of its parts, with the {\it more}, in the case of the singlet state, being encapusulated by the semi group $SM(4,C)$.  However, it should be possible to extend this approach, in general, to  GHZ states. This becomes even more important given the recent work on entangled qudits \cite{qudits}. The GHZ states must be invariant under some yet to be determined algebraic structure that cannot be reduced to the sum of its parts. For example, the GHZ state given by
\begin{equation}\Psi =\frac{1}{\sqrt 2}\biggl[\left(
                                  \begin{array}{c}
                                    1 \\
                                    0 \\
                                  \end{array}
                                \right)
\otimes \left(
                                  \begin{array}{c}
                                    1 \\
                                    0 \\
                                  \end{array}
                                \right)
\otimes \left( \begin{array}{c} 0\\ 1\\ \end{array}\right)\otimes \left( \begin{array}{c} 0\\ 1\\ \end{array}\right)
- \left(\begin{array}{c}
                                    0 \\
                                    1 \\
                                  \end{array}
                                \right)
\otimes \left( \begin{array}{c} 0\\1\\ \end{array}\right)\otimes \left(
                                  \begin{array}{c}
                                    1 \\
                                    0 \\
                                  \end{array}
                                \right)
\otimes \left(
                                  \begin{array}{c}
                                    1 \\
                                    0 \\
                                  \end{array}
                                \right) \biggr]\end{equation}
can in turn be written as $ (0,0,0,1,0,0,0,0,0,0,0,0,-1,0,0,0)^T$ and is  invariant under the $SM(16,C)$, where $a_{i4} =a_{i(13)}$, $i \ne 4, i\ne 13$ and $a_{44} +a_{4(13)} =1$ and 
$a_{(13)4}+a_{(13)(13)}=1$ .   

This article serves as a first step in classifying these inherent symmetries associated with perfect correlations and  has suggested a general approach to handling the statistics of multparticle systems. Our interpretation is based on a critical realist
approach which takes the given data to be objective, without falling
into some Newtonian dualistic interpretation of reality. We do not
put the observer outside of the world he or she  is observing, as is the
approach in classical mechanics, nor do we go to the other extreme of
giving a singular importance to the observer. We interact with the physical world because we are part of
it and indeed it is precisely this interaction that
permits us to do physics. The laws of interaction are part of the
objective laws of the universe and are there to be discovered.
Quantum physics has tried to incorporate those interactions
pertaining to physics into its axioms. To the extent that it has
succeeded, quantum mechanics gives a more real and objective picture
of physical reality, the uncertainty relations being a case in
point. However, its success does not preclude that a more intuitive explanation of quantum theory might enrich our understanding of non-locality but in a way that preserves causality and  thus avoiding interpretations based on a  ``spooky action at a distance.''  The combination of the principle that the whole is more than the sum of the parts, taken together with $SU(2)$ relativity is meant to be a first  step in that direction. 
   

\end{document}